# Blockchains for Spectrum Management in Wireless Networks: A Survey

Matthew K. Luka, Okpo U. Okereke, Elijah E. Omizegba and Ejike C. Anene

*Abstract*—Regulatory radio spectrum management is evolving from traditional static frequency allocation and assignment schemes towards dynamic spectrum management and access schemes. This evolution is necessitated by a number of factors including underutilization of licensed spectrum bands, changing market and technological developments and increased demand for spectrum for emerging applications in multimedia communications, internet-of-things and fifth generation (5G) wireless networks. In simple terms dynamic spectrum management involves allowing unlicensed users known as secondary users (SUs) to access the licensed spectrum of a licensed user also known as primary user (PU). This is primarily achieved using spectrum sharing schemes that leverage spectrum database and cognitive radio techniques. However, the use of spectrum database and cognitive radio techniques faces reliability, security and privacy concerns for spectrum sharing. There is also a need to support other requirements of dynamic spectrum management such as secondary spectrum trading market and dynamic spectrum access coordination. In this work, we review the use of blockchains for enabling spectrum sharing and other aspects of dynamic spectrum management. The review covers the use of blockchain to record spectrum management information such as spectrum sensing results and spectrum auction transactions in a secure manner. The article also covers the use of smart contracts to support complex service-level-agreements (SLAs) between network operators which is key to supporting a self-organized secondary spectrum sharing market and enforcement of regulatory policies. A taxonomy of the intersection between blockchain and various concepts of dynamic spectrum management is also provided.

## I. INTRODUCTION

Radio spectrum is a valuable and limited resource that is used for wireless applications including satellite telemetry, tracking, and control; television and radio broadcasts; aeronautical and maritime navigation, broadband internet connections and aviation communication systems. Access to broadband internet helps improve quality of lives, creates job and market opportunities and drives economic growth [1]. Wireless networks such as cellular and Wi-Fi networks are the most viable means of providing broadband internet connection especially in underserved communities and developing countries. As the shown in Fig.1, the bandwidth required per internet user has progressively grown due to increasing demand for mobile broadband internet and other emerging wireless applications. However, there is big divide between the available bandwidth for internet users in developing countries and developed countries due to lack of adequate infrastructure and unfavorable economics. Even in the most developed nations there are gaps in wireless coverage, base stations and access points become overloaded in busy areas, thereby driving up the price [2]. Thus, there is a need to maximize radio spectrum using various spectrum management regimes aimed at maximizing gains from the use of the available spectrum by enforcing efficient usage of the spectrum while minimizing interference among users [3]. Traditionally, exclusive licensing spectrum management model is used by national spectrum managers to provide interference protection guarantees and allow higher power output for licensed users. However, this spectrum management regime faces two fundamental limitations [4]. Firstly, significant parts of the licensed spectrum is underutilized. Secondly, this command-and-control regime of spectrum management is slow to changing market and technological developments. These challenges coupled with the ever increasing demand for spectrum probed by multimedia and emerging services and application such as internet-of-things (IoT), smart grids, emergence of 5G (fifth generation) wireless network etc. has informed the need for new spectrum management methods to satisfy emerging connectivity needs [5]. Under exclusive

M. K. Luka is with the Department of Electrical and Electronics Engineering, Modibbo Adama University of Technology, Yola, Nigeria.He is currently a PhD student at Abubakar Tafawa Balewa University, Bauchi, Nigeria, (email: matthewkl@mautech.edu.ng)

O. U. Okereke is a professor of communication Engineering with the Electrical and Electronic Engineering Department, Abubakar Tafawa Balewa University, Bauchi, Nigeria (e-mail: uokereke@gmail.com)

E. E. Omizegba is a professor of control and systems Engineering with the Electrical and Electronic Engineering Department, Abubakar Tafawa Balewa University, Bauchi, Nigeria., (e-mail: eeomizegba@atbu.edu.ng).

E. C. Anene is a professor of Machine Control with the Electrical and Electronic Engineering Department, Abubakar Tafawa Balewa University, Bauchi, Nigeria., (e-mail: eanene@atbu.edu.ng).

regulatory scheme, spectrum efficiency can be achieved through spectrum re-farming. An example of spectrum re-farming is the reallocation of spectrum obtained from the transition of analogue television service to digital terrestrial television for mobile services.

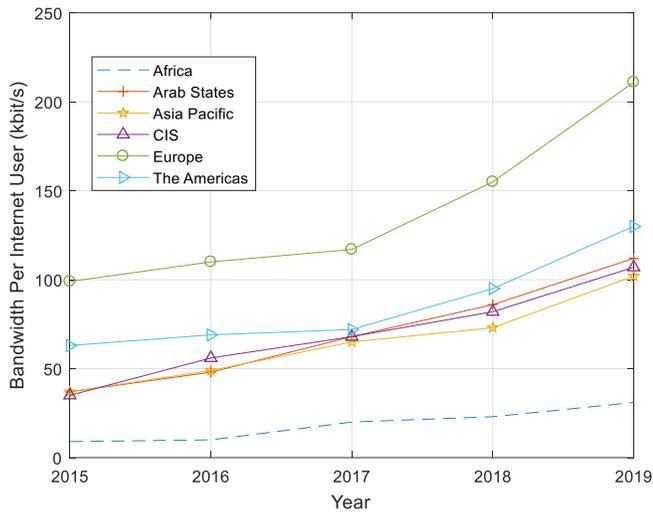

Fig.1 Bandwidth Demand per Internet User (Source: ITU World Telecommunication /ICT Indicators database)

A complimentary spectrum management tool is the use of license-exemption of spectrum bands such as 2.4 GHz ISM (Industrial Scientific and Medical) band to facilitate fast network deployment without incurring spectral license and associated costs. License-exempt spectrum management does not provide interference protection to radio services and its use is limited short-range devices (SRDs). Another complimentary method is to use frequency sharing to improve the utilization of existing spectrum resources. Spectrum sharing can be achieved at different levels [2]:

1) Between various radiocommunications applications or services in accordance with international, regional or national level regulatory arrangements;
2) Between various of users or entities (e.g commercial versus governmental);
3) Between licensed users of the same application (e.g mobile network operators);
4) Between licensed-exempt secondary users and protected primary users (e.g television station and regional broadband wireless network);
5) Between license-exempt users (e.g Wi-Fi and short range devices)

However, using spectrum sharing for dynamic spectrum management faces some challenges which are outlined in the next section. The rest of the article is organized as follows. Section II is a discussion on the challenges of various spectrum sharing schemes used for spectrum management. Section III provides an overview of the blockchain technology. In section IV, a comprehensive survey of various applications and uses of the blockchain technology for dynamic spectrum management is provided. Future research perspectives and challenges are discussed in section V. Finally, section VI concludes the article.

## II. CHALLENGES OF SPECTRUM SHARING SCHEMES

Dynamic spectrum sharing for the various spectrum management models can be achieved using cognitive radio technology. A cognitive radio system (CRS) uses technology that enables the system to derive knowledge of its geographical and operational environment, internal state and established policies to autonomously and dynamically adjust its operational protocols and parameters according to its obtained knowledge in order to learn from the results obtained and achieve predefined objectives [6]. The technical features that characterize a cognitive radio system are illustrated in Fig.2. The most popular means of facilitating dynamic spectrum sharing and access using CRS techniques in mobile and wireless network are spectrum sensing and geo-location with database access [7]. The salient aspects and challenges of spectrum sharing based on these techniques are outlined in this section.

### A. Spectrum Sensing Approach

Spectrum sensing is a cognitive radio technique used to detect spectrum holes or 'whitespace'. It is also used by secondary users (SU) to detect the arrival of primary users in the spectrum hole occupied by the SU. A taxonomy of various spectrum sensing techniques based on the size of the band of interest is presented in [8]. Narrowband spectrum sensing schemes decides whether a given portion of the spectrum is a whitespace or not while wideband sensing involves classifying individual channels of the spectrum to be either vacant or occupied. The main challenges of using spectrum sensing for spectrum sharing are outlined below [9]:

1) *Hidden Node Problem:*
   This problem occurs when a cognitive radio node cannot sense the presence of a receive-only node or another transmitting node (probably due to radio propagation conditions). When this happens, the cognitive radio node may assume that frequency channel is vacant.
2) *Sensing Capabilities:*
   spectrum sensing requires high sampling rate, sensitivity, high speed signal processors, high resolution and analogue-to-digital converters with large dynamic range. These requirements are difficult to fulfil with cognitive radio system devices that have constrained energy and limited processing capacity.
3) *Reliability of Sensing:*
   The reliability of spectrum sensing results is one of the main challenges impeding the widespread use of the technique for opportunistic spectrum access. The presence of external noise sources such as electric motors and AC power system can influence the reliability of spectrum sensing results.
   In addition to this key challenges, other issues associated with spectrum sensing are: sensing signaling cost; algorithm complexity with respect to power and processing demands; and cost of realizing cooperative sensing operations.

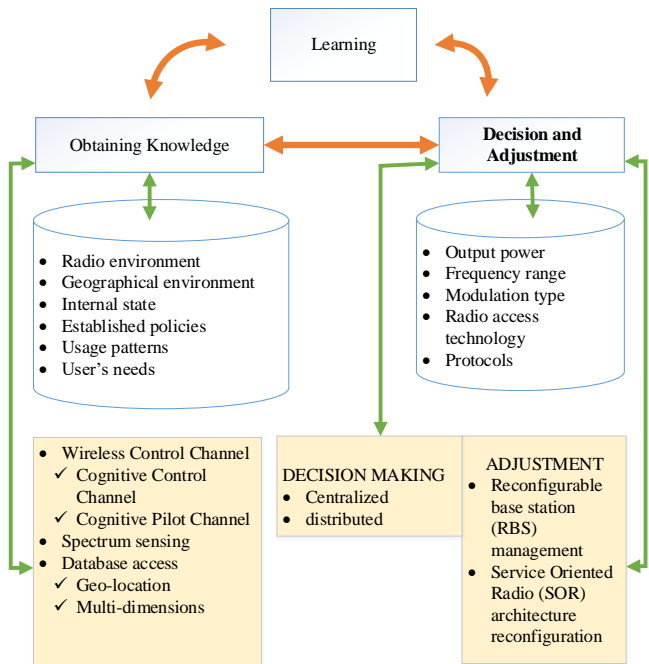

Fig.2. Illustration of Cognitive Radio System Concept [9]

## B. Database Approach

Spectrum database provides information about locally usable frequencies with the objective of providing protection to incumbent services from harmful interferences. The location of the CRS node in a database access scheme is determined using geo-location. Location information is needed to obtain correct information from the spectrum database for a given location. The database approach is mandatory in the TVWS technology and is the basis of both Licensed Shared Access (LSA) and Shared Access System (SAS) hierarchical spectrum sharing schemes. Some of these practical challenges as outlined in [10] include:

1) *Need for cross-national TVWS database carrier harmonization:*
   TVWS database should be capable of meeting the regulatory regime of different countries and regions while supporting interoperability to facilitate global and regional roaming of Television Whitespace Networks
2) *Need for a common inter-database communication protocol:*
   a particular country may have a number of TVWS database operators competing with each other in terms of network coverage, Quality of Service and broker fees. Thus there is a need for common communication protocol for TVWS database operators to coordinate TVWS information
3) *Lack of standardized validity period for database Queries:*
   Different spectrum operators specify different database query times.
4) *Establishment of a framework for contracting, authenticating and qualifying prospective database operators:*
   There is a need to outline clear-cut procedures to be adopted for qualifying a TVWS database operator
5) *Lack of standard procedure for framework testing:*

it is necessary to develop means of ensuring that TVWS database operators comply with regulatory, technical and operational procedures. It is also pertinent to verify that cognitive users comply with requirements for spectrum access

6) *Query language and format:*
   various TVWS database operators use different high-level language such as HTML, XML, Python and PHP. Therefore, there is a need to adopt a universal scripting language for uniformity and facilitating innovative ideas

Moreover, the use of database to manage fast varying and dynamic spectrum sharing can be challenging as the information stored in the repository can quickly become outdated [9]. The use of spectrum database also raises security and privacy concerns. Sensitive user information stored on the database must be protected from any unauthorized or unexpected access. Spectrum management can be seen as either a regulatory process or as a means of utilizing spectrum resources [11]. As a regulatory process, spectrum management can be defined as the implementation, coordination or realization of radio spectrum in order to meet the objectives of stakeholders. Based on this definition, spectrum management is process of rulemaking and implementing regulatory frameworks to achieve spectrum harmonization, spectrum allocation and other socio-economic benefits by regulatory authorities. Spectrum management can also be defined as controlling and making decisions on the utilization of spectrum resources to meet the objectives of various stakeholder [11]. This complementary definition reflects the evolution of spectrum management towards efficient utilization of spectrum as a resource, driven by technology and business use cases. Spectrum utilization is very crucial in the sub-6 GHz frequency band towards addressing the deficiency in spectrum requirements.

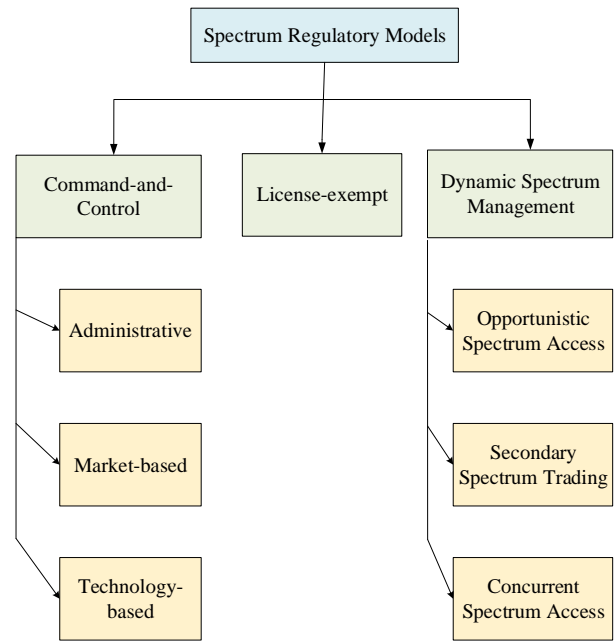

Fig.3 Spectrum Regulatory Models

The spectrum regulatory models for mobile and wireless communication can be categorized into three broad classes namely: Command and Control Model, Dynamic Spectrum Management Model and Spectrum Commons Model as illustrated in Fig.3. The various ways blockchain can be leveraged to provide a veritable platform for implementing the multi-faceted aspects of spectrum management are discussed in the next section.

## III. OVERVIEW OF BLOCKCHAIN TECHNOLOGY

Blockchain is an implementation of Distributed Ledger Technology (DLT) which enables large groups of nodes in the DLT network to reach consensus and record information without the need for a central authority. The main feature of Blockchain based DLT implementation that differentiates it from other DLT solutions is the storage of information in groups known as blocks, which are arranged as an ever growing chain of blocks with each validated block cryptographically linked to the previous block. A typical DLT system consist of DLT nodes, DLT user, service providers, and user group as illustrated in Fig. 4.

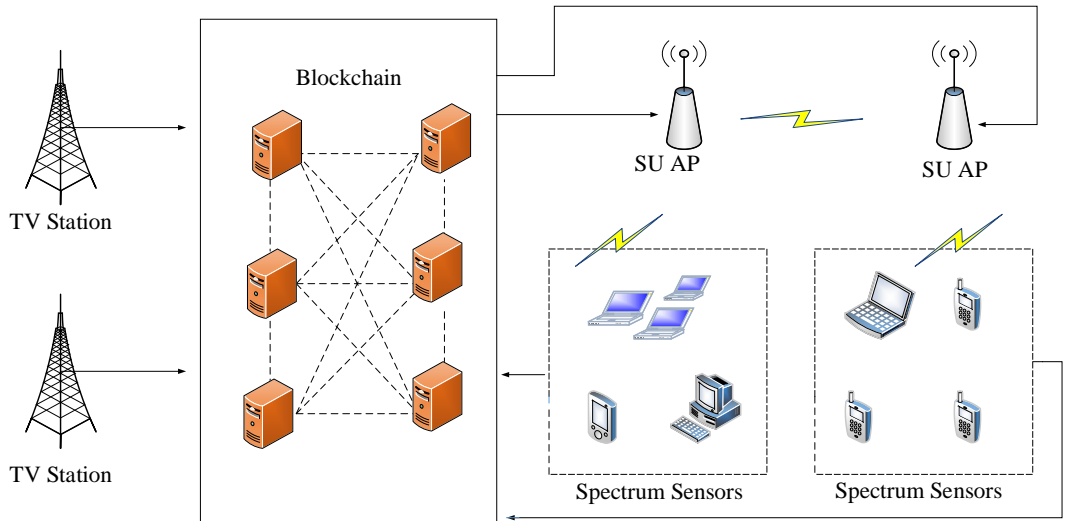

Fig. 4 DLT Actors and Components for use Case of Spectrum Sharing in TVWS

Nodes within the distributed ledger are responsible for storing ledger data, transferring the data to other nodes and validating newly added blocks. A service provider is a component of the DLT ecosystem that provides DLT based services to users through the interfaces it provides. Users consume services offered by the service provider. A certain user may be a service provider to another consumer. A user group is a collection of DLT users with a common feature. Blockchain networks can be broadly categorized as either *permissionless* (public) or *permissioned* (private or consortium). Permissionless blockchains are synonymous to the internet, where any user can participate and publish a new block without requiring permission from any authority. A user wishing to participate in a permissionless blockchain network can download any of the open-source blockchain software which are freely available online. Since anyone can read and write to the distributed ledger of a public blockchain, consensus algorithms are used to prevent malicious users from subverting the system. A number of consensus algorithms such as proof-of-work (PoW) and proof-of-stake (PoS) which requires users to maintain or expend resources for transactions are used to discourage malicious behavior in the network. Permissioned Blockchain network on the other hand is similar to an intranet network where only authorized users are allowed to update the network ledger or publish a new block. Permissioned network can either be private permissioned or consortium permissioned. A private permissioned Blockchain is limited to a single organization while a consortium permissioned Blockchain is made up of a number of decentralized organizations. The complexity of the consensus system used in a permissioned Blockchain network depends on the level of trust between participating users. Since the use of a computationally expensive algorithm may not be needed in a permissioned Blockchains, the transaction processing speed of these networks are faster than public blockchains. Different user access rights can be granted or revoked by the network authority of a permissioned Blockchain. The Key features of Blockchain which include decentralization, transparency, immutability, availability and security are useful for spectrum management in the following ways [13]:

1) *Decentralization*
    Implies that no third party is needed to validate transaction; users control their own data. This can potentially help to remove the need for database administrators

2) *Transparency*
    Since the ledger is shared among multiple nodes, the history of transaction and Blockchain algorithms are available for any component to review. This provides better

visibility for effective implementation of spectrum management rules.

3) *Immutability*

Blockchains are immutable and cryptographically secure, thus it is very hard to alter data on a Blockchain. This feature is useful in facilitating accurate auditing, enforcement and implementation of regulatory or contractual rules.

4) *Availability*

Blockchain ledgers are shared among nodes, thereby increasing the availability of the information stored on the blockchains. This makes Blockchain a more reliable spectrum management database for providing protection to incumbent users.

5) *Security*

All entries in the Blockchain ledger are cryptographically secured. Additionally all transactions between nodes are digitally signed and access to the blockchain by users is via a private/public cryptographic key pairs. This enhances the security of the wireless network infrastructure against attacks

In the next section, we will explore the use of blockchain technology as a platform for realizing a secure database scheme and implementing other aspect of spectrum sharing and dynamic spectrum management .

TABLE 1: APPLICATIONS OF BLOCKCHAIN FOR SPECTRUM MANAGEMENT IN MOBILE AND WIRELESS NETWORKS

| Application | Reference | Description | Other Aspects of Spectrum Management covered | | | |
|---|---|---|---|---|---|---|
| | | | SS | ST | SAC | SDE |
| Spectrum Sensing (SS) | [16] | Spectrum trading based on spectrum sensing and advertisement in ISM band | ✓ | ✓ | ✓ | x |
| | [17] | Cooperative Spectrum Sensing scheme for cognitive radio network | ✓ | x | ✓ | x |
| | [18] | Cooperative Spectrum Sensing scheme based on helper nodes for mobile networks | ✓ | x | ✓ | x |
| | [19] | Cooperative Spectrum Sensing scheme based on helper nodes for mobile networks | ✓ | x | ✓ | x |
| | [20] | Blochain based reputation score system for sensor nodes used for spectrum enforcement | ✓ | x | x | ✓ |
| Spectrum Trading (ST) | [21] | An auction based spectrum sharing scheme in which spectrum usage right is leased in exchange for digital currency | x | ✓ | ✓ | x |
| | [22] | Similar to the work in [21], with emphasis on the performance of the scheme under different fading channels | x | ✓ | ✓ | x |
| | [23] | Spectrum Trading between AP and UEs in a Blockchain-based RAN | x | ✓ | ✓ | x |
| | [24] | Spectrum trading using blockchain as a brokering system between WiFi access points | x | ✓ | ✓ | x |
| | [25] | Spectrum trading between MNOs using game theory | x | ✓ | ✓ | x |
| | [26] | Spectrum trading based on k-Vickrey auction mechanism in a spacecraft network | x | ✓ | ✓ | x |
| | [27] | Contract based spectrum sharing platform for M2M devices in a 5G network | x | ✓ | ✓ | x |
| | [28] | Spectrum trading using Stackelberg game theory between UAVs and MNOs | x | ✓ | ✓ | x |
| | [29] | Spectrum trading based on auction and free-market techniques for MNOs | x | ✓ | x | x |
| Spectrum Access Coordination (SAC) | [30] | Spectrum access coordination scheme based on consensus algorithm in a cognitive radio network | x | x | ✓ | x |
| | [31] | Spectrum access coordination modelled as a call admission process in a cognitive radio network | x | x | ✓ | x |
| | [32] | Spectrum access coordination based on Game theory in the unlicensed ISM band between MNOs | x | x | ✓ | x |
| | [33] | Spectrum access coordination based on Tit-for-Tat game theory for 5G-IoT devices | x | x | ✓ | x |
| | [34] | Spectrum access coordination for smallcells in a 5G network | x | x | ✓ | ✓ |
| | [35] | Spectrum access coordination in license-exempt spectrum using blockchain mining and spectrum auction | x | ✓ | ✓ | x |
| Secure Database and Policy Enforcement (SDE) | [36] | Blockchain platform for implements service level agreement between PWNOs and VMNO. Enforcement is ensured by the participation of a regulatory body. | x | ✓ | ✓ | ✓ |
| | [37] | BC was used to interface secondary users with SAS database in order to ensure privacy of users while enforcing SAS policies via smart contracts. | x | x | ✓ | ✓ |
| | [38] | Blockchain used as a secure platform to implement a learning-based MAC spectrum sharing protocol for battlefield equipment. | x | x | x | ✓ |
| | [39] | Use blockchain to manage service level agreements between MNOs in a software define network. | x | x | x | ✓ |

## IV. APPLICATIONS OF BLOCKCHAINS FOR SPECTRUM MANAGEMENT

Blockchain is new technology that could have a disruptive effect when used to replace existing solutions. Thus it is important to determine if a Blockchain is really needed for a particular application. In [14], the authors outlined a number of questions on a chart to determine if Blockchain technology can solve a particular problem. The following provides answers to the questions with respect to spectrum management:

1) Can *a traditional database meet your need?*

Traditional centralized database approach has been used for spectrum management in various access systems such as the Spectrum Access System, Licensed Spectrum Access and TVWS whitespace networks. Traditional geolocation database approaches cannot meet the needs of a self-organized spectrum market in which primary users can lease their spectrum for a fee. It cannot also enforce the complex contractual agreements and policy regulations associated with market driven spectrum management [15].

2) *Does more than one participant need to be able to update the data?*

Spectrum management involves multiple stakeholders such as regulatory agencies, primary users and secondary users. Secondary Users need to provide geo-location and sensing information for coordinated access to spectrum resources. Primary Users can advertise available frequencies for spectrum trading. Spectrum regulators may also need to audit spectrum usage or enforce regulatory policies.

3) *Do all those updaters trust one another?*

There is no trust among updaters; there exist a possibility for a malicious Secondary User to report false location information. A dishonest Primary User can also attempt to sell the same spectrum to two Secondary Users in the same location.

4) *Would all the participants trust a third party?*

Participants are unlikely to trust a third party due to the possibility of harvesting sensitive user data such as location for adversarial aims. Participants need to directly control the data to guarantee accuracy of information.

5) *Does the data need to be kept private?*

Spectrum management data should be shared among network participant but not necessarily with the general public.

6) *Do you need to control who makes changes to the Blockchain software?*

In order to address the challenge of governance, the spectrum regulatory body of a country can be saddled with the responsibility of managing the Blockchain

Based on this guide, a permissioned Blockchain network can be recommended for spectrum management in wireless and mobile networks. A literature review of recent research articles presented in Table 1, shows that both permissioned and permissionless Blockchains can be used for various aspects of spectrum management including spectrum sensing, spectrum sharing and policy enforcement. Blockchains are also amenable to other components of dynamic spectrum management such as realizing a geolocation database and facilitating secondary spectrum trading markets. This section discusses some of the applications of Blockchains for spectrum management in wireless networks

A. *Spectrum Sensing platform*

Blockchain can be leveraged to implement a secure spectrum sensing framework and to enable cooperative sensing, both of which enhances the reliability of spectrum sensing results. Using spectrum sensing allows MNOs to aggregate usable vacant frequencies with their licensed frequencies to achieve increased network capacity. The work in [16] leverages blockchain to implement a smart contract for spectrum sensing and spectrum trading. A SU first carries out spectrum sensing to identify a vacant channel. Once a vacant channel is identified, the SU invokes a smart contract function requesting for the vacant channel from the PU. If the PU agree to the price and time duration of lease proposed by the SU, the spectral usage right is leased in exchange for a native cryptocurrency. However, the PU may not always cooperate with SUs to lease its frequency on a contractual basis. In this case, SUs will seek to opportunistically access any unused spatiotemporal channel.

The reliability of spectrum sensing results can be improved through cooperative sensing, which involves fusing the sensing results from a number of SUs or sensors. This helps minimize the adverse effect of shadowing, multipath fading and noise uncertainty. In [17], a cooperative spectrum sensing scheme based on blockchain mobile networks was proposed. In this work, SUs can participate in the communication network as spectrum sensors and as mining nodes in the blockchain network. Each spectrum sensor broadcast its sensing result to other nodes on the blockchain network. The sensing results are then added as a transaction block on the blockchain through mining process. A user who wish to access a vacant channel invokes a smart contract that allows it bid for the vacant channel. The best bid for the vacant channel is selected by a module of the smart contract, which activates the payment of cryptocurrencies to miners and transfer of spectral usage right to the winner of the bid. However, both spectrum sensing and blockchain mining operations are energy intensive and computationally expensive tasks that may cause additional operational cost and latency.

The work in [19] is also a cooperative sensing scheme that is similar in concept to the work in [18] but with variations in the smart contract formulation and malicious helper nodes detection algorithm. The smart contract spectrum trading platform facilitates crowd-sourcing of spectrum sensing from a heterogeneous pool of helper nodes. Malicious nodes detection is achieved by the smart contract using a *cluster-based Helper Identification* algorithm. The sensing reports from helpers are compressed to minimize transaction cost of executing transaction on the blockchain. Honest helper nodes are paid after verification that they have provided spectrum service. Shifting the responsibility of spectrum sensing to helpers reduces the prohibitively high cost of deploying sensors by mobile network operators. The security concerns that may arise as a result of malicious or false helper nodes activities can be handled by smart contract algorithms running on the blockchain. The use of multiple helpers also improves the accuracy of the distributed cooperative spectrum sensing scheme enabled by blockchain. The work in [16] places the responsibility of coordinating spectrum access on the PU, while the works in [17], [18] and [19] allows SU to coordinate spectrum using cooperative spectrum sensing schemes implemented on blockchain. Distributed cooperative sensing implemented on blockchain can also be used by spectrum regulators to realize distributed enforcement of spectrum polices. The authors in [20] use blockchain to evaluate the trustworthiness of sensor nodes used in a spectrum enforcement system. Each sensor is assigned a reputation score based on its reported location and signal-to-noise ratio. The works uses the *most difficult chain* rule rather than the popular *longest chain* to resolve multiple forks for all nodes to converge at the same blockchain state

## B. Self-Organized Spectrum Trading Market

Implementation of secondary spectrum trading requires due consideration for techno-economic factors. Some of the technical solutions supported by TVWS networks include, network optimization, determination of accurate terrain data and channel propagation model as well as effective geo-location database design to realize practical deployment [10]. Although different economic models such as the auction model, contract model, spot market model and commodity model have been developed to augment the technical solutions, there are a number of economic challenges that impede real world roll out of the technology. Some of the economic challenges associated with database-assisted spectrum management approach include [10]:

1) The need for design specifications for currency and payment platforms
2) The need for universal standards, bidding language and economic models for the various database administrators and spectrum lessor and lessee.
3) Lack of specifications for contract enforcement modules to ensure primary user satisfaction while meeting secondary user's quality of service requirements.

The inherent characteristics and features of blockchain can be used to address these challenges associated with the economics of spectrum trading. Blockchain was originally implemented as a peer-to-peer payment platform. Thus it inherently lends itself as a solution for realizing a veritable spectrum payment system using digital currency that can be easily converted to fiat money. The various function of geo-location database and requirements of spectrum administrators can be implemented on blockchain platform. The bidding language, economic model, primary user protection requirements, QoS needs of users and other contract enforcement modules can be implemented on blockchain platforms using smart contracts.

Real-time auction mechanisms can be used for secondary spectrum markets using the blockchain framework as demonstrated in [21] and [22]. In both works, the primary user advertises its vacant channel using a smart contract deployed on a blockchain platform. Users who wish to access the available channel submit bids within a slotted time frame. The PU leases the spectrum usage right to the winning bid based on first-come-first-served basis. The exchange of spectrum usage right and payment for the lease is broadcast in the blockchain network. Secondary users with enough computational resources compete to mine the transaction using a variant of proof-of-work consensus algorithm and are compensated using a cryptocurrency which can be used to bid for spectrum in the next spectrum auction timeframe. The work in [22] differs from the work in [21] by taking into account the effect of multipath diversity on simulation results.

Spectrum sharing in heterogeneous spacecraft network belonging to different organization must meet trust and privacy concerns of the operators due to the sensitivity of their operations. In [26], the authors use the K-Vickrey auction mechanism to implement a smart contract application on a blockchain for a heterogeneous spacecraft network. The regulatory organization publishes a smart contract containing information about the available channels and the length of time-slot for spectrum access. The security of the auction mechanism is enhanced through the use of ElGamal homomorphic cryptosystem in the distributed ledger for both encryptions and decryptions of bid prices, bidders' addresses and clearing price.

Auctions mechanisms for spectrum trading can be regarded as a game with the objective of analyzing the dominant equilibrium between sellers and buyers [10]. The work in reference [25] use smart contact on a consortium blockchain to implement a spectrum trading spectrum sharing scheme based on game theory. PWNO compete with each other using the non-cooperative game of pricing to attract secondary mobile network operators. However, the amount of spectrum shared between the lessor and the lessee depends on parameters such as the number of connected end users, the modulation technology, the total spectrum owned by the PWNO and the spectrum utilization. The result of the game model gives a Nash Equilibrium, which is a combination of series of strategies such that the strategy of each player in the combination is optimal with respect to the strategies of other players. Similarly, the relationship between spectrum sellers and buyers was formulated as a Stackelberg game strategy in [28]. The Stackelberg game formulation consist of a leader (spectrum owner, in this case a MNO) and followers which are Unmanned Aerial Vehicles (UAV). The Stackelberg game is formulated to derive the optimal spectrum pricing and purchasing strategies to maximize the revenues of the UAV and MNO operators. The MNO advertises available spectrum based on its spectrum pricing policy using a smart contract on a consortium blockchain. Each UAV then responds with its spectrum purchasing policy based on the price given by the MNO. The smart contract then selects the winning bid by determining the Stackelberg Equilibrium point which meets profit objectives of both the MNO and UAVs.

In [27], a consortium blockchain is used to support spectrum trading between primary users and Machine-to-Machine (M2M) SUs. The base station (BS) which acts as the gateway for the M2M devices publishes a smart contract on the blockchain to motivate PUs with vacant spectrum resources for spectrum sharing. The contract specifies the amount of required spectrum and the corresponding reward. The contract between a PU and the BS gateway is added to the blockchain using Proof-of-work consensus algorithm. The spectrum shared by the PUs is allocated to SUs using a combinatorial one-to-one matching between PUs and SUs.

Blockchain is being considered as an enabling technology for providing decentralized security and efficient spectrum sharing for the sixth-generation (6G) wireless network [40]. The implementation of a blockchain radio access network (RAN) in [23] considers spectrum as the basic digital asset that is traded between the access point (AP) and user equipment (UE). Under this platform, spectrum asset represents a short-term right to exclusively transmit or receive with fixed power over a specific channel within a particular geographic area. Information about any spectrum asset is recorded on the blockchain. The AP allocates the asset based on current status of information stored on the blockchain. A cached smart contract concept is used for fast and paid access to the RAN. The work in [24] also proposes the use of spectrum brokering between WiFi APs to alleviate the problem of congestion in the ISM band. Although the work describes a centralized brokering platform, a distributed platform based on blockchain was proposed for future implementation to facilitate spectrum trading in the ISM band.

Blockchains provides an enhanced platform to support secondary spectrum trading market where a MNO can choose to be spectrum seller or spectrum buyer; depending on its spectrum demand during a time period. The work in [29] leverages a consortium blockchain to propose a spectrum trading platform between MNOs. In this framework, the spectrum demand of each MNO is predicted by its Operations Administration and Maintenance (OAM) servicer based on spectrum usage results collected from the operator's base stations. Spectrum buyers and sellers exchange spectrum usage rights using a double auction strategy which comprises of two phases. In the first phase, the selling price and buying prices advertised by sellers and buyers are matched and successful matches are added as transactions to the blockchain to complete transfer of spectral usage rights. In the second stage, unsuccessful auctions bids and asks are adjusted to realize additional transactions within a given time period.

### C. Spectrum Access Coordination

The interaction between primary users and secondary users or between secondary users in a dynamic spectrum management system can be modelled using machine learning models, biologically and socially inspired models or game theoretical models [41]. According to game theory, players makes decisions for different actions to attain various goals [42]. Game theory can be used as a mathematical framework for spectrum management and spectrum access coordination since the choice of each player determines the outcome of the game. Several researchers leverage blockchain to implement spectrum sharing coordination and cooperation schemes based on game theory. The Nash Equilibrium game theoretical technique proposed in [32] assumes that all Mobile Network Operators (MNO) participating in the spectrum sharing are allocated equal amount of cryptocurrency tokens to be used for accessing the unlicenced spectrum band. The token spent by a MNO for accessing the spectrum is shared equally among all other participating MNO. In order for the operator to maximize its use of its limited token, it has to balance its spectrum access strategy using the Nash Equilibrium function:

$$w_k = \frac{W}{n}, \quad k \in [1, n], \tag{1}$$

Where $k$ is the number of MNOs, $W$ is the total available bandwidth and $w_k$ is the amount of utilized spectrum. The work in [33] expresses spectrum sharing as a non-zero game in which the outcomes of gains and losses attained by each player is not zero; using a Tit-for-Tat (TFT) game theory that can be implemented as a smart contract on a blockchain. The TFT strategy allows spectrum users to cooperate first and then retaliate if other peers betray and cooperate again if other users cooperate. Thus the strategy seeks to combine the features of cooperative and non-cooperative game theory for internet-of-things (IoT) devices to participate in coordinated spectrum access.

Spectrum access coordination among secondary users can also be achieved using a consensus algorithm as shown in the consensus-before-talk (CBT) scheme in [30]. The CBT scheme aims to achieve a collision-free distributed spectrum access by exchanging spectrum access requests and reaching consensus on spectrum access using the distributed ledger. The CBT framework consists of a spectrum access transaction (SAT), a consensus policy module (CPM) and a distributed spectrum ledger (DSL). A SU's spectrum access request is embedded in a SAT and exchanged with all other SUs. The consensus protocol is applied to each SAT broadcast on the blockchain network. Once a SAT is validated, the result is stored in each SU's local DSL. The CPM implements the Byzantine Fault Tolerance (BFT) consensus algorithm to ensure liveliness, correctness and consistency of SATs. Before applying the consensus algorithm to new SATs, the SATs are stored in the SU's Spectrum Access Queue (SAQ) segment of the local ledger. After applying the consensus algorithm, the SATs are stored the SU's Spectrum Access History (SAH) segment of the DSL. SAT scheduling can be based on either a first-verified-first-served basis of SATs in SAQ or using a fairness scheme for least served SATs in the SAH.

Under the scheme proposed in [34], each small cell in a 5G network requests a particular amount of unlicensed spectrum for a specified amount of time. This request is broadcast to other small cells in a cluster of small cells on the blockchain network. Each member of the cluster votes to either grant or deny access to the requesting small cell. If majority of the cluster nodes vote to grant spectrum access to the small cell node, the node allows its user equipments (UEs) to transmit or receive on the new spectrum. The UEs then pay for spectrum access using cryptocurrency. Similar to the approach in [34], the work in [35] is also based on the election of a leader who becomes the exclusive owner of a portion of the spectrum for a period of time. The election process is achieved using PoW consensus protocol. Where more than one leader emerges from the PoW consensus protocol, the PoS consensus algorithm is used to select the overall winner. The winner can choose to either use the spectrum or lease the usage right to other users for a fee in an auction.

Another approach to spectrum access coordination is to model the process as a call admission control optimization problem using Continous Time Markov Decision Process (CTMDP) as proposed in [31]. Spectrum access requests of opportunistic secondary users are queued using a blockchain before they can access a vacant spectrum channel. The parameters of the CTMDP were applied to a feed-forward neural network to derive the optimal policy to maximize rewards of the threshold policy for accepting SUs.

### D. Secure Spectrum Sharing Database and Policy Enforcement

One of the challenges of using a centralized database system is the dynamic nature of wireless networks arising from SU mobility and variation of traffic demands of PUs. Using blockchain to dynamically record information of vacant spectrum bands and geo-location of users is envisioned to improve the efficiency of spectrum access and utilization. Blockchain can be used to record information regarding interference protection needs of PUs as well as spectrum usage and availability of TV whitespaces with respect to frequency, geo-location and time [15]. The use of blockchain allow users to control data in the ledger and ensure accuracy of information.

The reference in [37] seeks to address security and privacy concerns of SUs (second tier and third tier users) in a centralized spectrum access system (SAS). This is to prevent sensitive user data such as identities, spectrum usage, location, and transmission power from being exploited for adversarial purposes. The framework uses blockchain to interface General Authorized Access (GAA) to SAS database administrators so as to address privacy concerns while complying with regulatory design requirements. The smart contract is implemented as a distributed application that interfaces multiple database administrators to the regulatory body and GAA users. The GAA users are SUs that seek to opportunistically access spectrum whitespaces. The spectrum regulator registers both database administrators and GAA users who wish to participate in the spectrum sharing scheme. Registered GAA users form clusters and choose a cluster leader which anonymously authenticates with the geo-location databases through the blockchain platform. This is done to enhance scalability and privacy of cluster members. Each cluster leader queries the database after which members of the cluster reach a consensus on how the spectrum resources are to be shared among them. After a spectrum assignment consensus is reached, the cluster leader notifies the database administrators regarding the spectrum channels used by its members as required by spectrum regulation policy. The database administrator and other cluster leaders serve as blockchain nodes to reach a consensus on the validity of this information using Byzantine Fault Tolerance (BFT) consensus algorithm.

Applications of internet-of-thing in battlefield equipment used for military operations are expected to be robust against security threats such as denial of service attacks and data falsification. The work in [38] devised a blockchain based spectrum sharing protocol for secure operations of battlefield equipment against denial-of-service attacks. The denial-of-service attack is assumed to originate from an adversary equipment that generates false data in order to influence the spectrum access decision so as to degrade network performance. It is also assumed that the malicious equipment has large computation and storage resources and is capable of mimicking one of the equipment. This security threat is further mitigated by using a parameter known as Decision Parameter (DP) which is incorporated into a PoW-variant consensus protocol. The DP of each equipment is computed from the average number of back-offs per seconds experienced by other equipment that used the blocks it added for making spectrum access decision in previous timeframes. Using the DP improves the chances of user with good data to add blocks and penalizes users that have previously provided false spectrum sensing data. In addition to preventing malicious users from adding false data to the blockchain, every equipment that wants to transmit data is expected to use the historic data from the latest 7 blocks to make spectrum access decision using a learning based MAC protocol.

Blockchain can be used to implement secure service level agreements (SLAs) between Mobile Network Operators (MNOs) and Mobile Virtual Network Operators (MVNOs) These SLAs deployed as smart contacts can be used for sharing spectrum resources and network infrastructure between MNOs and MVNOs. A blockchain framework that brings together MNOs, MVNOs and regulatory bodies for spectrum management is proposed in [36]. The SLAs coded as smart contracts on the blockchain implements the regulatory policies of the regulatory body as well as the spectrum sharing conditions between MNOs and MVNOs. This ensures that subscribers of the participating MVNO are provided with better quality of service than subscribers of an opportunistic MVNO. The work in [36] also combines blockchain with software defined networking (SDN) to enhance spectrum management process and facilitate seamlessly roaming of subscribers from one MNO to another. SDN enable users to dynamically switch between MNOs whenever an operator's service is unavailable. Smart contracts used on the unified platform implement the spectrum sharing agreements and manage the complexities of SLAs among stakeholders. The smart contract also automates the process of billing and the task of roaming settlements between MNOs.

## V. FUTURE RESEARCH PERSPECTIVES AND CHALLENGES

The use of blockchains for spectrum management is an emerging application that has a lot of prospects and challenges. Some of the future research perspectives are discussed in this section.

### A. Large Scale spectrum Sensing Campaigns

TVWS Technology was originally designed to be driven by spectrum sensing algorithms [10]. However, results of occupancy measurements done with low cost devices such as Short Range Devices (SRDs) and Radio Frequency Identification (RFID) at a limited number of locations have shown that spectrum sensing cannot satisfactorily help identify possible bands that might be used for white space applications [7]. The use of cooperative spectrum sensing and the deployment of helper spectrum sensors have shown promise for use as part of a dynamic spectrum access. Thus, there exist an opportunity to leverage blockchain, cooperative spectrum sensing techniques and large scale deployment of fixed receiving stations and other mobile data harvesting techniques to carry out spectrum occupancy campaigns. This approach is likely to generate meaningful complimentary information to facilitate spectrum management.

### B. Combining Spectrum Sensing and Geo-Location Database

The two main methods of facilitating dynamic spectrum access are spectrum sensing and Geo-location database. Previous research works have considered this approaches as disparate techniques. Blockchain being a database technology can be leveraged to develop a unified approach where spectrum sensing techniques and geo-location database technology can be used as complimentary to each other. Combing these two spectrum access techniques will result in a more robust dynamic spectrum management framework.

### C. Decentralized Artificial Intelligence

Artificial Intelligence (AI) techniques such as statistical machine leaning, deep learning and deep reinforcement learning have been used for various aspects of spectrum management such as spectrum sensing [43] and dynamic spectrum access [44]. These AI algorithms work better when data is obtained from a database that is trusted, credible, reliable and secure [45]. Combining blockchain and AI creates a framework for Dynamic Spectrum Management that ensures secure data sharing, decentralized intelligent computing, explainable AI decisions, and coordinating untrusting devices [46]. Other benefits of this synergy include: secure and scalable blockchains, privacy-preserving personalization and automated governance and referee. Spectrum usage data stored on the blockchain can be used to achieve periodic re-training of the machine learning algorithms, thereby enhancing robustness to the dynamic wireless radio environment. Thus, there is an opportunity to consolidate blockchain and Artificial Intelligence to achieve superior spectrum management decision and spectrum access performance.

### D. Latency and Throughput Performance

The throughput of a blockchain framework measured in transactions per seconds (TPS) and latency (measured in seconds) are largely determined by the consensus algorithm used to validate transactions in the network. The architectural model of the blockchain network also plays a significant role in the performance of the network. Existing public blockchain networks such as Bitcoin and Ethereum tend to have very low throughput and high latency due to the complexity of consensus protocols used. Private and Consortium blockchains have much higher throughput and lower latency. Hyperledger consortium combines flexible consensus algorithms and an *execute-order-validate* architectural model to achieve performance throughput of up to 10,000 tps and a latency of around 0.5 s. However, further studies are required on the impact of blockchain performance on the different aspects dynamic spectrum management in wireless networks.

### E. Blockchain Network Deployment

There is a need to study the deployment of the blockchain with the communication network. The blockchain network can be deployed as an overlay of the communication network to allow nodes in the communication network act as full nodes on the blockchain network. However, this network setup can be energy consuming and requires a dedicated control channel for exchanging transactions and blocks over the blockchain networks [15]. Dependence on the control channel poses a risk to operation of the blockchain network due to susceptibility of the control channel to jamming by malicious users. An alternative network deployment approach is to use a dedicated blockchain platform so that the blockchain works as an independent database. While this approach is desirable for offloading the energy-intensive task of mining to the blockchain network, it deprives users of the opportunity to directly verify transactions. Thus, a third option is for communication nodes to participate in the blockchain network as user nodes to enable users participate in the transaction verification process but offload the mining task to full nodes of the blockchain network. Additional studies is needed to establish the relationship between these network deployment options and other factors such competition among secondary users, revenue maximization for communication network operators, throughput and latency performance etc. It is also important to investigate the performance of blockchain-based spectrum management system when combined with other emerging networking techniques such as mobile edge computing and software defined networking.

### F. CONCLUSION

In this work, a survey of recent literature on the use of blockchain for spectrum sharing and dynamic spectrum management is presented. Using blockchain offers a number of technical and business benefits over centralized geo-location database. Some of the technical benefits include: transparent ledger for spectrum auditing and management; high availability; strong security and use of smart contracts to handle complex sharing arrangements and enforcement of regulatory policies. The business benefits include: incentivizing users to share spectrum or perform spectrum sensing; users maintaining control over own data and removing the need for central database operator. The survey also provides a taxonomy of the various ways blockchain is used for spectrum management. Works of spectrum sensing, spectrum access coordination, secondary spectrum trading market and use of blockchain as a secure spectrum database were covered. The survey of literature shows that blockchain is a veritable platform for spectrum management.

### REFERENCES

[1] McKinsey, "Internet Matters: The Net's Sweeping Impact on Growth, Jobs and Prosperity," *Online and Upcoming: The Internet's Impact on Aspiring Countries*, 2011. [Online]. Available: http://www.mckinsey.com/insights/high_tech_telecoms_internet/internet_matters. [Accessed: 10-Mar-2020].

[2] ITU-R, "Draft Final Report for Resolution 9: Participation of Countries, Particularly Developing Countries, in Spectrum Management," 2017.

[3] M. Cave, C. Doyle, and W. Webb, *Essentials of Modern Spectrum Management*. Cambridge: Cambridge University Press, 2007.

[4] P. Anker, "From spectrum management to spectrum governance," *Telecomm. Policy*, no. May 2016, pp. 1–12, 2017.

[5] M. Massaro, "Next generation of radio spectrum management : Licensed shared access for 5G," *Telecomm. Policy*, no. November 2016, pp. 0–1, 2017.

[6] ITU-R, "Introduction to Cognitive Radio Systems in the Land Mobile Service." Geneva, pp. 1–17, 2011.

[7] ITU-R, "Spectrum Management Principles, Challenges and Issues Related to Dynamic Access to Frequency Bands by Means of Radio Systems Employing Cognitive Capabilities." Geneva, pp. 1–71, 2017.

[8] A. Ali and W. Hamouda, "Advances on Spectrum Sensing for Cognitive Radio Networks: Theory and Applications," *IEEE Commun. Surv. Tutorials*, vol. 19, no. 2, pp. 1277–1304, 2017.

[9] ITU-R, "Cognitive Radio Systems in the Land Mobile Service," vol.

1–71. ITU, Geneva, 2014.

[10] A. H. Kelechi, R. Nordin, and N. F. Abdullah, "Database-Assisted Television White Space Technology: Challenges, Trends and Future Research Directions," *IEEE Access*, vol. 4, pp. 8162–8183, 2016.

[11] IEEE Std 1900.1™-2019, "IEEE Standard for Definitions and Concepts for Dynamic Spectrum Access: Terminology Relating to Emerging Wireless Networks, System Functionality, and Spectrum Management." IEEE, 2019.

[12] ITU-T, "Distributed Ledger Technology Terms and Definitions." ITU-T, 2019.

[13] M. B. H. Weiss, K. Werbach, D. C. Sicker, and C. E. C. Bastidas, "On the Application of Blockchains to Spectrum Management," *IEEE Trans. Cogn. Commun. Netw.*, vol. 5, no. 2, pp. 193–205, 2019.

[14] M. E. Peck, "Do you need a Blockchain?," *IEEE Spectr.*, pp. 38–39, 2017.

[15] Y. Liang, *Dynamic Spectrum Management: From Cognitive Radio to Blockchain and Artificial Intelligence*. Singapore: Springer, 2020.

[16] T. Ariyarathna, P. Harankahadeniya, S. Isthikar, N. Pathirana, H.M. N. D. Bandara, and A. Madanayake, "Dynamic Spectrum Access via Smart Contracts on Blockchain," in *IEEE Wireless Communications and Networking Conference*, 2019, pp. 1–6.

[17] Y. Pei, S. Hu, F. Zhong, D. Niyato, and Y. Liang, "Blockchain-enabled Dynamic Spectrum Access: Cooperative Spectrum Sensing, Access and Mining," in *2019 IEEE Global Communications Conference (GLOBECOM)*, 2019, pp. 1–6.

[18] S. Bayhan, A. Zubow, and A. Wolisz, "Spass: Spectrum Sensing as a Service via Smart Contracts," in *IEEE International Symposium on Dynamic Spectrum Access Networks (DySPAN)*, 2018, pp. 1–10.

[19] S. Bayhan, A. Zubow, P. Gawłowicz, and A. Wolisz, "Smart contracts for spectrum sensing as a service," in *IEEE Transactions on Cognitive Communications and Networking*, 2019, pp. 1–13.

[20] M. A. A. Careem and A. Dutta, "SenseChain: Blockchain based Reputation System for Distributed Spectrum Enforcement," in *2019 IEEE International Symposium on Dynamic Spectrum Access Networks (DySPAN)*, 2019.

[21] K. Kotobi and S. G. Bilen, "Blockchain Enabled Spectrum Access in Cognitive Radio Network," in *2017 Wireless Telecommunications Symposium*, 2017.

[22] K. Kotobi and S. g. Bilén, "Secure Blockchains for Dynamic Spectrum Access," *IEEE Veh. Technol. Mag.*, vol. 13, no.1, pp.32–39, 2018.

[23] Y. Le, X. Ling, J. Wang, and Z. Ding, "Prototype Design and Test of Blockchain Radio Access Network," in *2019 IEEE International Conference on Communications Workshops*, 2019, pp. 1–6.

[24] F. Den Hartog, F. Bouhafs, and Q. Shi, "Toward secure trading of unlicensed spectrum in cyber-physical systems," in *16th IEEE Annual Consumer Communications & Networking Conference*, 2019.

[25] S. Han and X. Zhu, "Blockchain Based Spectrum Sharing Algorithm," in *IEEE 19th International Conference on Communication Technology*, 2019, pp. 936–940.

[26] L. Yu, J. Ji, Y. Guo, Q. Wang, T. Ji, and P. Li, "Smart Communications in Heterogeneous Spacecraft Networks: A Blockchain Based Secure Auction Approach," in *IEEE Cognitive Communications for Aerospace Applications Workshop*, 2019, pp. 1–4.

[27] Z. Zhou, X. Chen, Y. Zhang, and S. Mumtaz, "Blockchain-Empowered Secure Spectrum Sharing for 5G Heterogeneous Networks," *IEEE Netw.*, vol. 34, no. 1, pp. 24–31, 2020.

[28] J. Qiu, D. Grace, and G. Ding, "Blockchain-Based Secure Spectrum Trading for Unmanned Aerial Vehicle Assisted Cellular Networks: An Operator's Perspective," *IEEE Internet Things J.*, vol. 7, no.1, pp. 451–466, 2020.

[29] S. Zheng, T. Han, Y. Jiang, and X. Ge, "Smart Contract-Based Spectrum Sharing Transactions for Multi-Operators Wireless Communication Networks," *IEEE Access*, vol. 8, pp. 88547–88557, 2020.

[30] H. Seo, J. Park, M. Bennis, and W. Choi, "Consensus-Before-Talk: Distributed Dynamic Spectrum Access via Distributed Spectrum Ledger Technology," in *IEEE International Symposium on Dynamic Spectrum Access Networks (DySPAN)*, 2018, pp. 1–7.

[31] W. Ni, Y. Zhang, and W. Li, "Optimal Admission Control For Secondary Users using Blockchain Technology In Cognitive Radio Networks," in *IEEE 39th International Conference on Distributed Computing Systems*, 2019, pp. 1518–1526.

[32] T. Maksymyuk, J. Gazda, L. Han, and M. Jo, "Blockchain-Based Intelligent Network Management for 5G and Beyond," in *3rd International Conference on Advanced Information and Communications Technologies*, 2019, pp. 36–39.

[33] Y. Choi and I. Lee, "Game Theoretical Approach of Blockchain-based Spectrum Sharing for 5G-enabled IoTs in Dense Networks," in *IEEE 90th Vehicular Technology Conference (VTC2019-Fall)*, 2019, pp. 1–6.

[34] V. Sevindik, "Autonomous 5G Smallcell Network Deployment and Optimization in Unlicensed Spectrum," in *IEEE 2nd 5G World Forum*, 2019, pp. 446–451.

[35] X. Fan and Y. Huo, "Blockchain Based Dynamic Spectrum Access of Non-Real-Time Data in Cyber-Physical-Social Systems," *IEEE Access*, vol. 8, pp. 64486–64498, 2020.

[36] D. B. Rawat and A. Alshaikhi, "Leveraging Distributed Blockchain-based Scheme for Wireless Network Virtualization with Security and QoS Constraints," in *International Conference on Computing, Networking and Communications*, 2018, pp. 332–336.

[37] M. Grissa, A. A. Yavuz, and B. Hamdaoui, "TrustSAS: A Trustworthy Spectrum Access System for the 3.5 GHz CBRS Band," in *IEEE Conference on Computer Communications*, 2019, pp. 1495–1503.

[38] M. Patnaik, G. Prabhu, C. Rebeiro, V. Matyas, and K. Veezhinathan, "ProBLeSS: A Proactive Blockchain based Spectrum Sharing Protocol against SSDF Attacks in Cognitive Radio IoBT Networks," *IEEE Netw. Lett.*, vol. 2, no. 2, pp. 67–70, 2020.

[39]  A. A. Okon, I. Elgendi, O. S. Sholiyi, J. M. H. Elmirghani, A. Jamalipour, and K. Munasinghe, "Blockchain and SDN Architecture for Spectrum Management in Cellular Networks," *IEEE Access*, vol. 8, pp. 94415–94428, 2020.

[40]  T. Huang, W. Yang, J. Wu, J. Ma, X. Zhang, and D. Zhang, "A Survey on Green 6G Network : Architecture and Technologies," *IEEE Access*, vol. 7, pp. 175758–175768, 2019.

[41]  M. A. Shattal, A. Wisniewska, A. Al-fuqaha, S. Member, B. Khan, and K. Dombrowski, "Evolutionary Game Theory Perspective on Dynamic Spectrum Access Etiquette," *IEEE ACCESS*, vol.6, pp.1–15, 2017.

[42]  A. S. M. Z. Shifat and M. Z. Chowdhury, "Game Theory Based Spectrum Sharing for QoS Provisioning in Heterogeneous Networks," in *2016 5th International Conference on Informatics, Electronics and Vision (ICIEV)*, 2016, pp. 272–275.

[43]  K. M. Thilina, K. W. Choi, N. Saquib, and E. Hossain, "Machine Learning Techniques for Cooperative Spectrum Sensing in Cognitive Radio Networks," *IEEE J. Sel. Areas Commun.*, vol. 31, no.11, pp. 2209–2221, 2013.

[44]  N. Zhao, Y.-C. Liang, D. Niyato, Y. Pei, and Y. Jiang, "Deep Reinforcement Learning for User Association and Resource Allocation in Heterogeneous Networks," in *EEE Global Communications Conference (GLOBECOM'18)*, 2018, pp. 1–6.

[45]  K. Salah, M. H. Rheman, N. Nizamuddin, and A. Al-Fuqaha, "Blockchain for AI: Review and Open Research Challenges," *IEEE Access*, vol. 4, pp. 1–23, 2018.

[46]  T. N. Dinh and M. T. Thai, "AI and Blockchain: A Disruptive Integration," *Cimputer*, pp. 48–53, 2018.